\documentclass{article}
\usepackage{arxiv}

\usepackage[utf8]{inputenc} 
\usepackage[T1]{fontenc}    
\usepackage{url}            
\usepackage{booktabs}       
\usepackage{amsfonts}       
\usepackage{nicefrac}       
\usepackage{microtype}      
\usepackage{graphicx}
\usepackage{natbib}
\usepackage{doi}
\usepackage{listings}
\usepackage{float}

\title{What do people want to fact-check?}

\author{
Bijean Ghafouri\thanks{Corresponding author: bghafour@usc.edu} \\
University of Southern California \\
Los Angeles, CA, USA \\
\And
Dorsaf Sallami \\
Mila \\
Montreal, QC, Canada \\
\And
Luca Luceri \\
University of Southern California \\
Los Angeles, CA, USA \\
\And
Taylor Lynn Curtis \\
Mila \\
Montreal, QC, Canada \\
\And
Jean-Francois Godbout \\
Mila \& Université de Montréal \\
Montreal, QC, Canada \\
\And
Emilio Ferrara \\
University of Southern California \\
Los Angeles, CA, USA \\
\And
Reihaneh Rabbany \\
Mila \& McGill University \\
Montreal, QC, Canada \\
}

\renewcommand{\shorttitle}{What do people want to fact-check?}

\hypersetup{
  pdftitle={What do people want to fact-check?},
  pdfsubject={cs.CY, cs.AI},
  pdfauthor={Bijean Ghafouri, Dorsaf Sallami, Luca Luceri, Taylor Lynn Curtis, Jean-Francois Godbout, Emilio Ferrara, Reihaneh Rabbany},
  pdfkeywords={Fact-checking demand, Fake-news, Misinformation, User-generated claims, AI verification systems},
  hidelinks
}

\begin{document}
\maketitle

\begin{abstract}
Research on misinformation has focused almost exclusively on supply, asking what falsehoods circulate, who produces them, and whether corrections work. A basic demand-side question remains unanswered. When ordinary people can fact-check anything they want, what do they actually ask about? We provide the first large-scale evidence on this question by analyzing close to 2{,}500 statements submitted by 457 participants to an open-ended AI fact-checking system. Each claim is classified along five semantic dimensions (domain, epistemic form, verifiability, target entity, and temporal reference), producing a behavioral map of public verification demand. Three findings stand out. First, users range widely across topics but default to a narrow epistemic repertoire, overwhelmingly submitting simple descriptive claims about present-day observables. Second, roughly one in four requests concerns statements that cannot be empirically resolved, including moral judgments, speculative predictions, and subjective evaluations, revealing a systematic mismatch between what users seek from fact-checking tools and what such tools can deliver. Third, comparison with the FEVER benchmark dataset exposes sharp structural divergences across all five dimensions, indicating that standard evaluation corpora encode a synthetic claim environment that does not resemble real-world verification needs. These results reframe fact-checking as a demand-driven problem and identify where current AI systems and benchmarks are misaligned with the uncertainty people actually experience.
\end{abstract}


\newpage
\section{Introduction}

We know a great deal about the supply side of misinformation. Researchers have documented its prevalence across digital platforms \citep{lazer2018science, vosoughi2018spread, grinberg2019fake, guess2020exposure}, identified the cognitive mechanisms that make it persuasive \citep{pennycook2019lazy, pennycook2021shifting, lewandowsky2021countering, ecker2022psychological}, and shown that corrections can reduce misperceptions \citep{lewandowsky2021countering, roozenbeek2022psychological}. Fact-checking infrastructures now operate at global scale \citep{pennycook2020fighting, guo2022survey, aimeur2023fake, sallami2025exploring}. Yet this large body of work leaves a deceptively simple question unanswered. When people encounter uncertainty in everyday life, what do they actually want to verify?

The question matters because the supply of checkable claims and the demand for verification need not coincide. Existing fact-checking pipelines prioritize claims that are salient, viral, or selected by professional editors \citep{micallef2022true, sehat2024misinformation, cazzamatta2025truth, graves2016deciding}. But the claims that generate genuine uncertainty for individuals may look very different. User-initiated fact-checking is a form of epistemic outsourcing \citep{amazeen2015revisiting, graves2017anatomy}, in which people delegate the work of verification to experts, institutions, or machines \citep{hardwig1985epistemic, goldman2001experts, zagzebski2012epistemic}. Mapping this behavior reveals where users perceive gaps between what they know and what they can confirm. It also exposes how ordinary people draw boundaries between fact, inference, and value, boundaries that often diverge from scientific or philosophical standards \citep{waisbord2018truth, starmans2012folk, macfarlane2014assessment, elgin2017true}. Without demand-side evidence, we cannot design AI systems that match real cognitive needs, and we cannot evaluate whether benchmark datasets reflect the verification problems users actually face.

Existing datasets used to train and evaluate fact-checking models do not fill this gap. They are constructed by journalists, researchers, and professional fact-checkers, and they capture expert judgments about what should be checked rather than the actual distribution of uncertainty experienced by users \citep{graves2016deciding, wang2017liar, thorne2018fever}. Models trained on these corpora inherit this skew and may generalize poorly when deployed on real user queries that contain a very different mix of claim types. Community-driven moderation systems face a complementary problem. Platforms that rely on user-generated notes depend entirely on contributor activity \citep{chuai2024did}, meaning that claims attracting little attention may never be evaluated, leaving large segments of real verification demand unaddressed.

This paper provides the first large-scale description of what users choose to fact-check in an open-ended AI verification system. We analyze close to 2,500 statements submitted by 457 participants to an interactive fact-checking platform \citep{curtis2025veracity} and classify each claim along five semantic dimensions that capture domain, epistemic form, verifiability, target entity, and temporal orientation. Three main findings emerge. First, users range widely across topics but default to narrow epistemic habits, overwhelmingly submitting descriptive and directly verifiable claims even as their topical interests vary. Second, roughly one in four requests targets statements that cannot be empirically resolved, including moral judgments and speculative predictions, revealing a mismatch between what users seek from fact-checking and what such systems can deliver. Third, comparing these distributions with FEVER \citep{thorne2018fever}, a widely used benchmark, exposes sharp structural divergences across all five dimensions, indicating that standard evaluation corpora encode a claim environment that does not resemble real-world verification demand.

Our results reframe fact-checking as a demand-driven problem. They complement existing research on misinformation supply and correction effects by documenting where individuals seek epistemic closure. They also provide a behavioral baseline for the design of next-generation verification tools and for understanding where current models and benchmarks are structurally misaligned with the uncertainty people actually experience.

\section{Methodology}

\subsection{Data Collection}

We recruited 1{,}025 participants via \textit{Prolific} and directed them to an interactive fact-checking application~\citep{curtis2025veracity}. Within the application, participants submitted statements they were personally uncertain about. 

To ensure data quality and assess participant attentiveness, an instructional manipulation check was included. Participants were asked to disregard the statement content and instead select both the “\textit{Strongly Agree}” and “\textit{Strongly Disagree}” response options. Responses from participants who failed the instructional manipulation check were 
excluded from all subsequent analyses. The final sample comprised 457 participants. Each respondent provided multiple claims, averaging five per user, yielding 2,473 total claims, all in English.

We also collected self-reported demographic information from participants. Among those who provided complete data, the sample was predominantly between 25 and 54 years old (approximately 77\%), with smaller groups of younger and older adults. A slight majority identified as female (55.58\%), and most participants reported at least some college education.  Participants spanned the political spectrum in both ideology and party identification. Detailed distributions for age, gender, education, employment, income, political ideology, and party identification are reported in Appendix~\ref{app:demographics}.

\subsection{Data Processing}
We used GPT-OSS-20B~\citep{agarwal2025gpt} in an instruction-following setup to automatically assign a single label for each taxonomy dimension 
(topic, epistemic form, empirical verifiability, target entity, and time frame). This taxonomy captures the semantic structure required to distinguish the 
content of a claim, its epistemic framing, and the types of evidence that 
could potentially verify it. 
Model outputs were generated using a fixed prompt (see Appendix~\ref{app:prompts}) and 
temperature $t = 0.0$, ensuring deterministic label assignments. 

Each claim includes a veracity score from zero to one that reflects the strength of the evidence found for the statement~\citep{curtis2025veracity}. A score of zero indicates that the claim is unsupported or false, a score of one indicates that the claim is well supported or true, and a score near zero point five indicates mixed or uncertain evidence. The score is generated by a large language model that evaluates the claim in conjunction with retrieved web evidence. 

\section{Results}
The sections below summarize each taxonomy dimension and report the distribution of labels across the dataset.
\subsection{Domain of Claims}
To measure what kinds of topics users chose to fact-check, we applied a domain classifier that assigns each statement to one of seven areas. The scheme distinguishes among \textit{Politics and Government}, \textit{Economy and Business}, \textit{Health and Medicine}, \textit{Science and Technology}, \textit{Environment and Climate}, \textit{Society and Culture}, and \textit{Lifestyle or Miscellaneous} topics. These seven areas reflect standard fact-checking domains and capture the broad topical space in which users typically seek verification.
As shown in Table~\ref{tab:domains}, users most often fact-check Science and Technology (22.3\%) or Politics and Government (18.6\%), with lower but still a substantial interest in Lifestyle, Health, Society, Environment, and Economy, and with veracity scores that vary only modestly across domains.

\begin{table}[htbp]
\centering
\caption{Distribution of domain classifications\vspace{-10pt}
}
\label{tab:domains}
\begin{tabular}{lcc}
\toprule
Label & Proportion (\%) & Veracity \\
\midrule
Science/Technology & 22.3 & 0.62 \\
Politics/Government & 18.6 & 0.51 \\
Lifestyle/Misc & 16.1 & 0.49 \\
Health/Medicine & 13.9 & 0.55 \\
Society/Culture & 12.8 & 0.58 \\
Environment/Climate & 8.5 & 0.62 \\
Economy/Business & 7.7 & 0.56 \\
\bottomrule
\end{tabular}
\end{table}

\subsection{Epistemic Type}
To analyze how users articulate truth claims beyond topical content, we applied an epistemic structure classifier that identifies the form of reasoning embedded in each statement. Each claim is assigned to one of four categories. Descriptive statements assert that something exists, occurred, or can be observed without numerical, causal, or evaluative content. Statistical statements quantify or compare phenomena with proportions, averages, or percentages. Causal statements posit relationships of dependency, including conditional or counterfactual structures. Normative statements express judgments or prescriptions such as moral claims or assertions about what should happen. This dimension captures how users frame knowledge rather than what the statement is about. These four categories isolate the fundamental reasoning structures through which people formulate truth claims, allowing us to distinguish whether a statement reports, quantifies, explains, or evaluates.

Table \ref{tab:epi} shows that users overwhelmingly frame their fact-checking requests as simple descriptive claims, while causal, statistical, and especially normative assertions appear less often.
\vspace{-5pt}
\begin{table}[htbp]
\centering
\caption{Distribution of epistemic classifications\vspace{-10pt}
} \label{tab:epi}
\begin{tabular}{lcc}
\toprule
Label & Proportion (\%) & Veracity \\
\midrule
Descriptive & 72.1 & 0.57 \\
Causal & 11.8 & 0.56 \\
Statistical & 11.5 & 0.51 \\
Normative & 4.5 & 0.45 \\
\bottomrule
\end{tabular}
\vspace{-10pt}
\end{table}

\subsection{Verifiability Level}
The third dimension captures how verifiable a claim is and whether its truth can be established through empirical evidence or only inferred. Verifiability identifies the kind of evidence required to adjudicate that knowledge. This approximates the practical boundary between observable fact, expert inference, and speculation. Each claim is labeled as Direct, Indirect, or Unverifiable. Direct claims can be validated through public records, official data, or direct observation, such as population counts or enacted laws. Indirect claims require expert interpretation, modeling, or inference. Unverifiable claims cannot be resolved empirically. They include moral evaluations or speculative predictions. These three categories capture the essential epistemic boundary between what can be observed, what must be inferred, and what cannot be empirically resolved. 

\begin{table}[htbp]
\centering
\caption{Distribution of verifiability classifications\vspace{-10pt}
}
\label{tab:veri}
\begin{tabular}{lcc}
\toprule
Label & Proportion (\%) & Veracity \\
\midrule
Direct & 64.3 & 0.61 \\
Unverifiable & 21.4 & 0.40 \\
Indirect & 14.3 & 0.56 \\
\bottomrule
 \end{tabular}
 \end{table}

The distribution in Table \ref{tab:veri} indicates that most users seek verification for observable facts while a substantial minority attempt to verify statements that lie outside empirical adjudication.

\subsection{Target Entity}
The fourth dimension identifies the referent of each claim: the entity, actor, or concept about which a factual assertion is made. The target classification captures who or what the claim concerns. Each statement is assigned one of six labels: \textit{individual}, \textit{organization}, \textit{group}, \textit{event}, \textit{policy}, or \textit{phenomenon}. Individuals denote named persons or roles tied to specific people. Organizations include corporate, governmental, or media institutions. Groups refer to demographic or social categories defined by shared traits. Events cover discrete or time-bounded occurrences such as elections or natural disasters. Policies represent laws, regulations, or formal government programs. Phenomena capture natural, conceptual, or mechanistic entities without agency, such as celestial bodies, environmental systems, or abstractions. These six categories capture the ontology of entities that claims can be about, distinguishing claims in a way that exhausts the space of plausible referents.

\begin{table}[htbp]
\centering
\caption{Distribution of target classifications\vspace{-10pt}
}\label{tab:targ}
\begin{tabular}{lcc}
\toprule
Label & Proportion (\%) & Veracity \\
\midrule
Phenomenon & 50.5 & 0.58 \\
Individual & 19.2 & 0.51 \\
Group & 15.8 & 0.58 \\
Organization & 7.5 & 0.56 \\
Event & 4.7 & 0.47 \\
Policy & 2.3 & 0.68 \\
\bottomrule
 \end{tabular}
 \end{table}

Table \ref{tab:targ} shows that most fact-checking requests concern phenomena rather than agents, suggesting that users often seek verification for claims about systems, mechanisms, and abstract processes. Claims about individuals, groups, and organizations appear less frequently, while events and policies account for a small share.

\subsection{Temporal Orientation}
The final dimension captures the time frame referenced by each claim. This dimension distinguishes whether statements describe past events, current conditions, future predictions, or timeless generalizations. Each claim is labeled as \textit{past}, \textit{present}, \textit{future}, or \textit{general}. Past statements reference completed events. Present statements describe ongoing conditions. Future statements express predictions or possibilities. General statements articulate enduring regularities that transcend specific time periods.

\begin{table}[htbp]
\centering
\caption{Distribution of temporal classifications\vspace{-10pt}
}
\label{tab:temp}
\begin{tabular}{lcc}
\toprule
Label & Proportion (\%) & Veracity \\
\midrule
Present & 42.8 & 0.55 \\
General & 29.9 & 0.61 \\
Past & 18.6 & 0.54 \\
Future & 8.7 & 0.46 \\
\bottomrule
 \end{tabular}
 \end{table}

The distributions in Table \ref{tab:temp} indicate that most fact-checking requests concern present conditions, with a substantial share targeting claims about enduring patterns. Past events appear less frequently, and yet future-oriented predictions comprise a non-negligible minority.

\subsection{User-Level Diversity}
To assess how broadly or narrowly individuals engage with different types of claims, we compute a Shannon entropy score for each user $i$ within each classification dimension $d$. For a dimension with $K_d$ categories, user-level entropy is:
\begin{equation}
H_{i}^{(d)} = - \sum_{k=1}^{K_d} p_{ik}^{(d)} \log p_{ik}^{(d)},
\end{equation}
where $p_{ik}^{(d)}$ is the proportion of claims from user $i$ that fall into category $k$. Higher values indicate that a user distributes their fact-checking across many categories (greater diversity), and lower values indicate concentration in a small subset (narrow focus). This metric provides a behavioral proxy for cognitive breadth.

\begin{table}[htbp]
\centering
\caption{User-level entropy across classification dimensions. Higher values indicate greater diversity in fact-checking behavior within that dimension.
}
\label{tab:entropy}
\begin{tabular}{lcc}
\toprule
\textbf{Dimension} & \textbf{Mean Entropy} & \textbf{SD} \\
\midrule
Domain        & 1.000 & 0.347 \\
Target        & 0.793 & 0.365 \\
Temporal      & 0.776 & 0.312 \\
Verifiability & 0.527 & 0.364 \\
Epistemic     & 0.495 & 0.360 \\
\bottomrule
\end{tabular}
\end{table}

\textit{Domain} shows the highest diversity ($H=1.00$), indicating that users range widely across issue areas rather than confining their attention to a single topical domain. In contrast, \textit{Epistemic Structure} ($H=0.49$) and \textit{Verifiability} ($H=0.53$) show the lowest entropy, suggesting that users tend to fact-check the same kinds of epistemic and verifiable forms. \textit{Target} and \textit{Temporal Orientation} exhibit mid-range diversity.  Overall, users display broad topical curiosity but narrow epistemic habits. They fact-check many different subjects, yet they do so within stable epistemic and evidentiary boundaries.

\subsection{Associations Between Semantic Dimensions}
Having established the distributions within each dimension, we next examine how these semantic properties co-occur.

Pairwise cross-tabulations reveal a small set of dominant co-occurrence patterns. Across domains, society/culture and lifestyle statements are almost entirely descriptive (84–90\%), while science/technology and economy/business contain the highest shares of statistical or causal language. Science/technology and environment/climate overwhelmingly target phenomena (79–84\%), and most statements in these domains are present-oriented (60–64\%).

Epistemically, statistical claims are almost always directly verifiable (85\%), causal claims are primarily indirect (47\%), and normative claims are largely unverifiable (63\%). Both causal and statistical statements overwhelmingly concern phenomena (64–66\%). Descriptive claims cluster in the present (57\%), whereas normative claims cluster in the general/timeless category (55\%).

Verifiability patterns further consolidate this tendency. Indirectly verifiable claims center on phenomena (76\%), and directly verifiable claims cluster in the present (55\%). Claims about organizations and policies are also strongly present-oriented (57–68\%). These patterns show that users structure claims in highly regular ways, aligning topical content, epistemic form, and evidentiary expectations. Full heatmaps association plots are reported in Appendix~\ref{app:cross}.

\subsection{Comparison with the FEVER Benchmark}
To assess how representative existing misinformation benchmarks are of real user fact-checking behavior, we applied our full classification pipeline to a random sample of 2,500 claims drawn from the FEVER, a widely used fact-checking dataset \citep{thorne2018fever}. The same five classifiers (\textit{Domain}, \textit{Epistemic Structure}, \textit{Verifiability}, \textit{Target}, and \textit{Temporal Orientation}) were used on this sample, allowing for a direct comparison between the distributions of both datasets. Benchmark datasets such as FEVER are designed for model training and evaluation, not to mirror real-world fact-checking demand. If the semantic distribution of FEVER diverges from user-generated data, systems trained on such benchmarks may be optimized for a task environment that differs from genuine public epistemic behavior. We provide the detailed comparison within each dimension in Appendix~\ref{app:fever}.

\begin{table}[htbp]\vspace{-10pt}
\caption{Average absolute difference (in percentage points) between classification distributions in the user dataset and FEVER dataset.\vspace{-10pt}
}
\label{tab:diff}
\centering
\begin{tabular}[t]{ccccc}
\toprule
Domain & Epistemic & Verifiability & Target & Temporal \\
\midrule
15.44 & 11.89 & 15.59 & 13.13 & 14.8 \\
\bottomrule
\end{tabular}
\vspace{-5pt}
\end{table}

Table \ref{tab:diff} exposes a sharp structural difference between benchmark data and real user fact-checking demand. FEVER’s topical distribution is dominated by entertainment-style statements (63\%), whereas these account for only 16\% of user queries. Users submit far more science and technology claims (22\% vs. 3\%), health and medicine (14\% vs. 2\%), and politics and government (19\% vs. 7\%).

Epistemically, FEVER is almost entirely descriptive (96\%), while the user dataset contains substantial causal, statistical, and normative content (a combined 28\%). FEVER claims are overwhelmingly directly verifiable (96\%), compared to 64\% in our data, and user-submitted claims are far more often unverifiable (21\% vs. 3\%). Target structure diverges as well: FEVER centers on individuals (54\%), whereas users focus on phenomena (51\%). Temporally, FEVER skews past-tense (56\%), while user queries leans toward present and general statements (43\% and 30\%).

We then compared the average veracity score in each dataset, a continuous measure from 0 (false) to 1 (true) reflecting how definitively a claim can be resolved, and find that user-submitted claims cluster much closer to the ambiguous middle (mean = 0.58, SD = 0.38) than FEVER’s sharply polarized claims (mean = 0.38, SD = 0.47), showing that real users most often seek verification for grey-area statements rather than the clearly true or false claims that dominate benchmark datasets.
These discrepancies indicate that benchmark corpora like FEVER capture only a narrow slice of the semantic space users bring to fact-checking systems, and models trained exclusively on such data are optimized for a claim environment that does not resemble real-world verification demand.

\section{Discussion}

Two central patterns emerge from our analysis. First, users overwhelmingly seek verification for simple, descriptive claims about observable present-day realities. Despite ranging widely across topical domains, they converge on a narrow set of epistemic forms: factual assertions that can be checked against public records or direct observation. This pattern suggests that most fact-checking behavior functions as \textit{cognitive anchoring}---users turn to verification systems not to reason through complex causal chains or evaluate statistical evidence, but to confirm or disconfirm concrete factual propositions they feel uncertain about. Uncertainty, as revealed by these data, is experienced primarily as a discrete informational gap rather than a broad epistemic crisis.

Second, a substantial share of queries (21.4\%) concern claims that cannot be empirically resolved, including moral evaluations, speculative predictions, and subjective judgments. Users attempt to outsource normative and predictive uncertainty to systems designed exclusively for empirical adjudication. This mismatch exposes a gap in public understanding of what fact-checking can accomplish and points to an unmet need: next-generation verification tools should not only assess evidence but also detect and communicate when a claim falls outside the scope of empirical resolution. Without this capacity, systems risk issuing spurious verdicts on inherently unverifiable statements, potentially reinforcing rather than correcting misplaced certainty.

These findings also reveal a structural limitation of existing benchmark datasets. The semantic profile of user-generated claims diverges sharply from FEVER across all five dimensions. FEVER is dominated by entertainment-related, past-tense, descriptive claims about named individuals---a profile that reflects its Wikipedia-derived construction rather than genuine public uncertainty. By contrast, users submit claims that are topically broader, epistemically more diverse, more often present-oriented, and frequently concern abstract phenomena rather than specific people. Models trained exclusively on benchmarks like FEVER are therefore optimized for a claim distribution that users rarely generate, which may partly explain poor generalization in deployed fact-checking systems. These results underscore the need for \textit{demand-driven} evaluation corpora that reflect the actual structure of public verification needs.

Several limitations qualify these conclusions. The sample consists of English-speaking Prolific users interacting with a single experimental application, which limits generalizability to other linguistic, cultural, and platform contexts. The taxonomy relies entirely on automated classification via a single language model, without human annotation for validation; future work should combine model-based labels with expert or crowd-sourced annotations to assess classification reliability. Additionally, the veracity scores used throughout the analysis are themselves model-generated, introducing the possibility that systematic biases in the scoring model affect the observed distributional patterns. Replication across different fact-checking platforms, languages, and annotator configurations would strengthen the conclusions drawn here.

Despite these limitations, the findings carry clear implications. Demand-side evidence should inform the design of both fact-checking tools and the benchmarks used to evaluate them. Mapping what users actually seek to verify provides a behavioral baseline for understanding how the public draws boundaries between fact, inference, and value---boundaries that shape expectations of AI systems and determine which forms of uncertainty people attempt to outsource. Building verification infrastructures that can navigate empirical adjudication and epistemic triage, distinguishing what can be checked from what cannot, remains a central challenge for the field.

\bibliographystyle{ACM-Reference-Format}
\bibliography{references}

\newpage
\appendix
\section{Demographic Characteristics}
\label{app:demographics}
\begin{table}[H]
    \centering
    \caption{Age distribution of participants}
    \label{tab:age}
    \begin{tabular}{lrr}
        \toprule
        Category & Count & Percent (\%) \\
        \midrule
        18--24   & 18  & 3.94  \\
        25--34   & 120 & 26.26 \\
        35--44   & 127 & 27.79 \\
        45--54   & 91  & 19.91 \\
        55--64   & 55  & 12.04 \\
        65+      & 23  & 5.03 \\
        Missing  & 23   & 5.03 \\
        \bottomrule
    \end{tabular}
\end{table}
\begin{table}[H]
    \centering
    \caption{Gender distribution of participants}
    \label{tab:gender}
    \begin{tabular}{lrr}
        \toprule
        Category & Count & Percent (\%) \\
        \midrule
        Female                      & 244 & 53.39 \\
        Male                        & 187 & 40.92\\
        Non-binary / third gender   & 5   & 1.09 \\
        Prefer not to say           & 3   & 0.66 \\
        Missing & 18 & 3.94  \\
        \bottomrule
    \end{tabular}
\end{table}
\begin{table}[H]
    \centering
    \caption{Education level distribution}
    \label{tab:education}
    \begin{tabular}{lrr}
        \toprule
        Category & Count & Percent (\%) \\
        \midrule
        Bachelor's degree (e.g., BA, BS)                          & 170 & 37.20 \\
        Some college, no degree                                   & 81  & 17.72 \\
        Master's degree (e.g., MA, MS, MEd)                       & 62  & 13.57 \\
        Associate degree (e.g., AA, AS)                           & 61  & 13.35 \\
        High school diploma or equivalent (e.g., GED)             & 43  & 9.41  \\
        Doctorate (e.g., PhD, EdD)                                & 16  & 3.50  \\
        Less than high school diploma                             & 3   & 0.66  \\
        Prefer not to say                                         & 2   & 0.44  \\
        Missing                                                   & 19   & 4.16  \\
        \bottomrule
    \end{tabular}
\end{table}
\begin{table}[H]
    \centering
    \caption{Employment status distribution}
    \label{tab:employment}
    \begin{tabular}{lrr}
        \toprule
        Category & Count & Percent (\%) \\
        \midrule
        Employed full-time               & 222 & 48.58 \\
        Self-employed                    & 54  & 11.82 \\
        Employed part-time               & 48  & 10.50 \\
        Unemployed, looking for work     & 47  & 10.28 \\
        Unemployed, not looking for work & 22  & 4.81  \\
        Retired                          & 21  & 4.60  \\
        Unable to work                   & 12  & 2.63  \\
        Prefer not to say                & 7   & 1.53  \\
        Student                          & 6   & 1.31  \\
        Missing & 18 & 3.94 \\
        \bottomrule
    \end{tabular}
\end{table}
\begin{table}[H]
    \centering
    \caption{Household income distribution}
    \label{tab:income}
    \begin{tabular}{lrr}
        \toprule
        Category & Count & Percent (\%) \\
        \midrule
        Less than \$25{,}000              & 96 & 21.01 \\
        \$25{,}000 to \$49{,}999          & 102 & 22.32 \\
        \$50{,}000 to \$74{,}999          & 77 & 16.85 \\
        \$75{,}000 to \$99{,}999          & 80 & 17.51 \\
        \$100{,}000 to \$149{,}999        & 44 & 9.63 \\
        \$150{,}000 or more               & 30 & 6.56  \\
        Prefer not to say                 & 9  & 1.97 \\
        Missing                           & 19  & 4.16  \\
        \bottomrule
    \end{tabular}
\end{table}
\begin{table}[H]
    \centering
    \caption{Political ideology distribution}
    \label{tab:ideology}
    \begin{tabular}{lrr}
        \toprule
        Category & Count & Percent (\%) \\
        \midrule
        Liberal               & 100 & 21.88 \\
        Moderate              & 88  & 19.26 \\
        Conservative          & 85  & 18.60 \\
        Extremely Liberal     & 54  & 11.82 \\
        Slightly Liberal      & 46  & 10.07 \\
        Slightly Conservative & 33  & 7.22  \\
        Extremely Conservative& 23  & 5.03  \\
        Other                 & 6   & 1.33  \\
        Prefer not to say     & 3   & 0.66  \\
        Missing               & 19   & 4.16  \\
        \bottomrule
    \end{tabular}
\end{table}

\begin{table}[H]
    \centering
    \caption{Party identification distribution}
    \label{tab:party}
    \begin{tabular}{lrr}
        \toprule
        Category & Count & Percent (\%) \\
        \midrule
        Independent                      & 118 & 25.82 \\
        Strong Democrat                  & 101 & 22.10 \\
        Strong Republican                & 73  & 16.63 \\
        Weak/Not very strong Democrat    & 71  & 15.97 \\
        Weak/Not very strong Republican  & 61  & 15.54 \\
        Other                            & 10  & 2.19  \\
        Prefer not to say                & 3   & 0.66  \\
        Missing                          & 20   & 4.38  \\
        \bottomrule
    \end{tabular}
\end{table}

\section{Prompt Templates}
\label{app:prompts}
This appendix presents the prompts used to elicit
domain, epistemic type, verifiability, target, and temporal labels from the language model classifier. For reproducibility, we report the full text of each prompt as implemented.

\subsection{Domain Classification Prompt}

\begin{lstlisting}
PROMPT_TEMPLATE = r"""
You are an expert domain annotator. Assign ONE domain label to the following claim.
Return ONLY valid JSON with the exact fields:
{"domain_label": "...", "confidence": 0-1, "rationale": "..."}


Fixed label set (choose exactly one):

1. Politics_Government-elections, politicians, laws, regulations, party behavior, corruption,
   public spending, and judicial or administrative acts.
   Verify via: legal statutes, government records, or official reports.
  Exclude if primarily about economic performance (-> Economy_Business) or medical mechanisms (-> Health_Medicine).


2. Economy_Business-markets, inflation, trade, GDP, prices, jobs, corporate actions, profits,
   layoffs, taxes, and financial assets.
   Verify via: financial or economic data.
   Exclude if government legislation or regulation is the core mechanism (-> Politics_Government).

3. Health_Medicine-diseases, treatments, vaccines, risks, nutrition, mental or physical health,
   medical research, clinical outcomes, or public health data.
   Verify via: clinical, biomedical, or epidemiological evidence.
   Exclude if about health policy (-> Politics_Government) or fundamental biological mechanisms
   (-> Science_Technology).

4. Science_Technology-scientific research, mechanisms, discoveries, engineering, AI systems,
   algorithms, data privacy, space exploration, and platform mechanics.
   Verify via: scientific or technical documentation.
   Exclude if focused on government regulation (-> Politics_Government), health outcomes
   (-> Health_Medicine), or economic consequences (-> Economy_Business).

5. Environment_Climate-global warming, pollution, weather, natural resources, biodiversity,
   and conservation.
   Verify via: environmental measurements or ecological reports.
   Exclude if mechanistic explanation (-> Science_Technology) or policy intervention
   (-> Politics_Government).

6. Society_Culture-identity, demographics, education, religion, beliefs, norms, immigration
   as social phenomenon, family behavior, and collective values.
   Verify via: surveys, cultural statistics, or sociological data.
   Exclude if artistic, entertainment, or consumer-product related (-> Lifestyle_Misc).

7. Lifestyle_Misc-entertainment, sports, celebrities, consumer products, travel, trivia,
   or speculative claims (fictional, science-fiction, conspiracy, unverified rumors).
   Verify via: product specifications, entertainment sources, or none (if unverifiable).
   Use only when no other domain clearly applies.

Decision rules:
- Choose based on the type of evidence that would confirm or falsify the claim:
   - Government statute or record -> Politics_Government
   - Market or financial data -> Economy_Business
   - Clinical or epidemiological study -> Health_Medicine
   - Scientific or technical documentation -> Science_Technology
   - Environmental measurement -> Environment_Climate
   - Survey or cultural statistic -> Society_Culture
   - Product spec, entertainment media, or unverifiable source -> Lifestyle_Misc

- Priority rules when ambiguous:
   - If both government and economy: prefer Politics_Government.
   - If both health and science: prefer Health_Medicine (applied mechanism over pure theory).
   - If both science and environment: prefer Science_Technology when explaining why;
    Environment_Climate when describing what.
   - If claim is speculative, conspiratorial, or entertainment-like: assign Lifestyle_Misc.

- Never infer or imagine context beyond the literal text.

Examples:
"The Senate passed a bill increasing farm subsidies" -> Politics_Government
"Inflation rose to 6 percent last quarter" -> Economy_Business
"The new diabetes drug reduces mortality" -> Health_Medicine
"This AI system passed the medical exam" -> Science_Technology
"Global temperatures hit record highs" -> Environment_Climate
"Most ninth graders do two hours of homework nightly" -> Society_Culture
"This streaming series is the best show" -> Lifestyle_Misc

Claim: "{claim}"
"""
\end{lstlisting}

\subsection{Epistemic Type Prompt}

\begin{lstlisting}
PROMPT_TEMPLATE = r"""
You are an expert annotator. Assign ONE epistemic label to the claim.
Return ONLY a single JSON object with keys:
{"epistemic_label": "...", "confidence": 0.0, "rationale": "..."}

Label set (choose exactly one):
- Descriptive
- Statistical
- Causal
- Normative

Definitions:
- Descriptive: states that something exists, is/was the case, or can be observed or recorded.
  Cues: "is/was/are," descriptive adjectives, historical or factual assertions, predictions that
  simply describe expected states without specifying causes.
  Examples: "The U.S. invaded Iraq in 2003." "Kamala Harris will run for president in 2028."

- Statistical: expresses quantity, rate, proportion, or measurable comparison without explicit
  causation.
  Cues: %, "half of," "increased by X," "doubled," "fewer than," "most people," "ratio," "average."
  Examples: "Unemployment is at 4.2%." "Most Americans own a smartphone."

- Causal: asserts cause/effect or dependency, including counterfactuals or conditional predictions.
  Cues: causes, leads to, due to, results in, increases the risk, "if... then," "as a result,"
  "because."
  Examples: "Smoking causes cancer." "Taking certain medications increases risk of autism."

- Normative: expresses value judgment, prescription, or moral evaluation.
  Cues: should/ought, right/wrong, good/bad, ethical/unethical, unconstitutional, best/worst.
  Examples: "People should recycle." "SORA is unconstitutional."

Decision boundaries:
1) If explicit number, proportion, or comparison and no causal language -> Statistical.
2) If causal or conditional language appears -> Causal (even if numbers are present).
3) If explicit value or moral judgment language appears -> Normative (even if facts or numbers
   are present).
4) If none of the above -> Descriptive.
5) If predictive ("will," "is expected to") but non-causal -> Descriptive, not Causal.
6) If moral tone is implied but not explicit -> Descriptive (avoid over-labeling as Normative).

Focus on epistemic *form* (how truth is presented), not on topic or accuracy.

Claim: "{claim}"
Output JSON only:
"""
\end{verbatim}

\subsection{Verifiability Prompt}

\begin{verbatim}
PROMPT_TEMPLATE = r"""
You are an expert annotator. Assign ONE verifiability label to the claim.
Return ONLY a single JSON object with keys:
{"verifiability_label": "...", "confidence": 0.0, "rationale": "..."}

Label set (choose exactly one):
- Direct
- Indirect
- Unverifiable

Definitions:
- Direct: truth can be checked using empirical data, official records, or direct observation.
  Examples: "California passed a new minimum wage law." "The population of Texas is 30 million."
  Criteria: measurable or observable facts; verifiable through public data or documentation.

- Indirect: requires expert interpretation, modeling, or inference; not directly measurable.
  Examples: "California passed a new minimum wage law." "The population of Texas is 30 million."

  Criteria: truth depends on models, correlations, or expert consensus.

- Unverifiable: cannot be confirmed or falsified using empirical evidence.
  Examples: "AI will destroy human creativity." "It's morally wrong to tax the rich."
  "JD Vance will run for President in 2028." "Donald Trump is on the Epstein list."
  Criteria: moral, metaphysical, speculative, or too far in the future to check.
  Includes subjective or taste-based judgments ("good," "bad," "top show"),
  and any prediction or conspiracy without empirical grounding.

Decision rules:
1) If verifiable via direct data, record, or observation -> Direct.
2) If requires interpretation or domain expertise -> Indirect.
3) If speculative, normative, or unfalsifiable -> Unverifiable.
4) If uncertain between Indirect and Unverifiable, choose Unverifiable unless
   the claim clearly references measurable processes or expert studies.
5) If the claim predicts a future event without current empirical basis -> Unverifiable.
6) If the claim expresses a subjective or aesthetic evaluation -> Unverifiable.

Do not explain topic content; assess only how checkable the claim's truth is.

Claim: "{claim}"
Output JSON only:
"""
\end{lstlisting}

\subsection{Target Classification Prompt}

\begin{lstlisting}
PROMPT_TEMPLATE = r"""
You are an expert annotator. Assign ONE target entity label to the claim.
Return ONLY a single JSON object with keys:
{"target_label": "...", "confidence": 0.0, "rationale": "..."}

Label set (choose exactly one):
- Individual
- Organization
- Group
- Event
- Policy
- Phenomenon

Definitions:
- Individual: a single named person or specific role referring to a person.
  Examples: "Biden said inflation is over." "Elon Musk tweeted about AI."
  Criteria: proper names, singular pronouns ("he," "she"), or roles tied to individuals.

- Organization: corporation, government agency, NGO, media outlet, or other institutional body.
  Examples: "The WHO approved the new vaccine." "Google launched Bard."
  Criteria: named collective entities that act as formal institutions.

- Group: demographic, social, or biological category of people or organisms.
  Examples: "Immigrants take American jobs." "Dogs are loyal animals."
  Criteria: plural or generic classes of beings defined by shared traits (age, species, ideology,
  nationality).

- Event: discrete, bounded, or time-linked happenings-political, natural, or social.
  Examples: "The 2020 election decided the outcome." "The earthquake killed 2,000 people."
  Criteria: specific occurrences, disasters, attacks, or named historical events.

- Policy: law, regulation, government program, or formal proposal.
  Examples: "The Green New Deal bans airplanes." "The Affordable Care Act expanded coverage."
  Criteria: codified rules or proposals within political or institutional contexts.

- Phenomenon: natural, celestial, or conceptual entity not covered above.
  Examples: "Venus rotates slowly." "Climate change is accelerating." "The Great Barrier Reef
  is bleaching."
  Criteria: physical objects, natural systems, or abstract concepts without agency or
  institutional structure.

Decision rules:
1) If the claim centers on a person -> Individual.
2) If a corporate, government, or institutional actor -> Organization.
3) If a collective of people or organisms -> Group.
4) If a bounded occurrence (past, present, or future) -> Event.
5) If about a law, bill, or program -> Policy.
6) If about a natural, celestial, or conceptual entity -> Phenomenon.
7) When uncertain between Group and Phenomenon, prefer Phenomenon for nonhuman subjects.

Focus on the main referent of the claim, not side mentions.
Claim: "{claim}"
Output JSON only:
"""
\end{lstlisting}

\subsection{Temporal Orientation Prompt}

\begin{lstlisting}
PROMPT_TEMPLATE = r"""
You are an expert annotator. Assign ONE temporal orientation label to the claim.
Return ONLY a single JSON object with keys:
{"temporal_label": "...", "confidence": 0.0, "rationale": "..."}

Label set (choose exactly one):
- Past
- Present
- Future
- General

Definitions:
- Past: refers to completed events or states.
  Cues: "was," "were," "had," "in 2019," "last year," "previous administration," "since."
  Example: "The U.S. invaded Iraq in 2003." "The reef has lost 50% of its coral since 1995."

- Present: describes ongoing or current conditions.
  Cues: "is," "are," "still," "currently," "now," "today," or simple present tense describing
  existing states.
  Example: "The economy is in recession." "There are 6.4 million people living in Atlanta."

- Future: predicts or projects outcomes that have not yet occurred.
  Cues: "will," "shall," "expected to," "going to," "set to," "predicted," "soon," or modal
  verbs like "might," "may," "could" when expressing possibility.
  Example: "AI will replace many human jobs." "This could cause prices to rise."

- General: expresses enduring or timeless truths without a specific temporal anchor.
  Cues: generic present tense with no time references, universal statements, proverbs,
  or scientific laws.
  Example: "Women earn less than men." "Taxes reduce economic efficiency."

Decision rules:
1) If explicit past time marker or past tense -> Past.
2) If present tense describing a current state or condition -> Present.
3) If predictive or modal ("will," "would," "might," "could," "expected") -> Future.
4) If statement is timeless or universal with no specific time reference -> General.
5) When uncertain, prefer Present over General unless the statement is explicitly universal.
6) If using present perfect ("has/have + past participle") describing completed change -> Past.

Focus only on temporal reference, not truth or sentiment.

Claim: "{claim}"
Output JSON only:
"""
\end{lstlisting}

\section{Cross-tab associations between classifications}
\label{app:cross}
\begin{figure}[H]
\centering
\begin{minipage}{0.45\linewidth}
  \includegraphics[width=\linewidth]{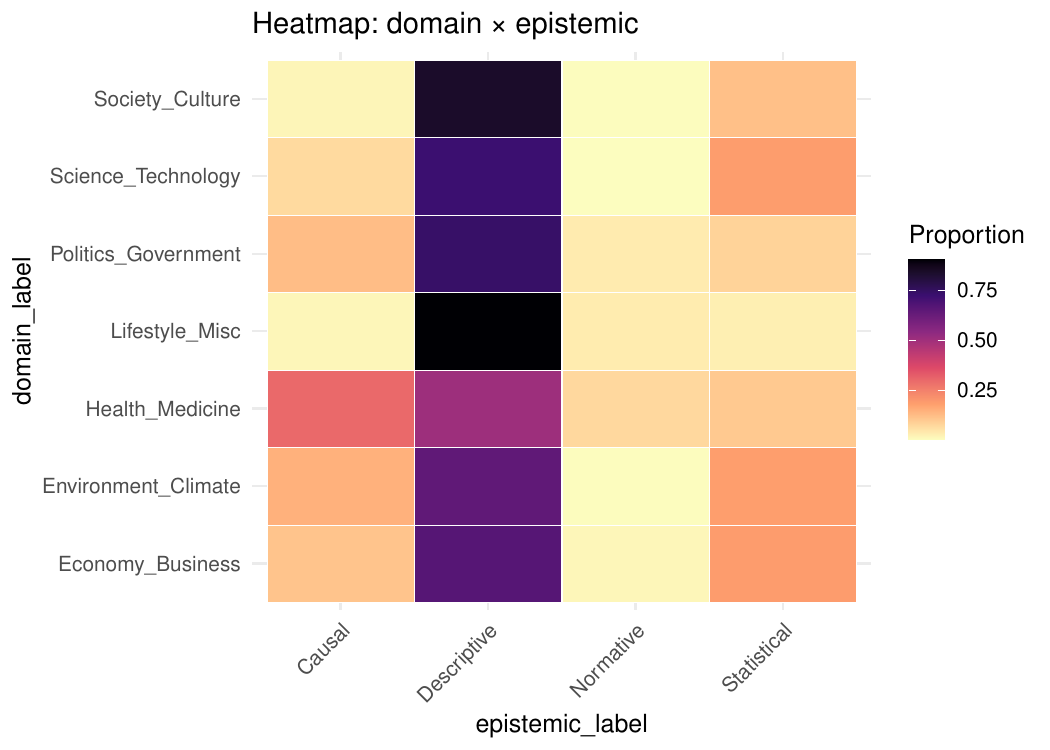}
  \includegraphics[width=\linewidth]{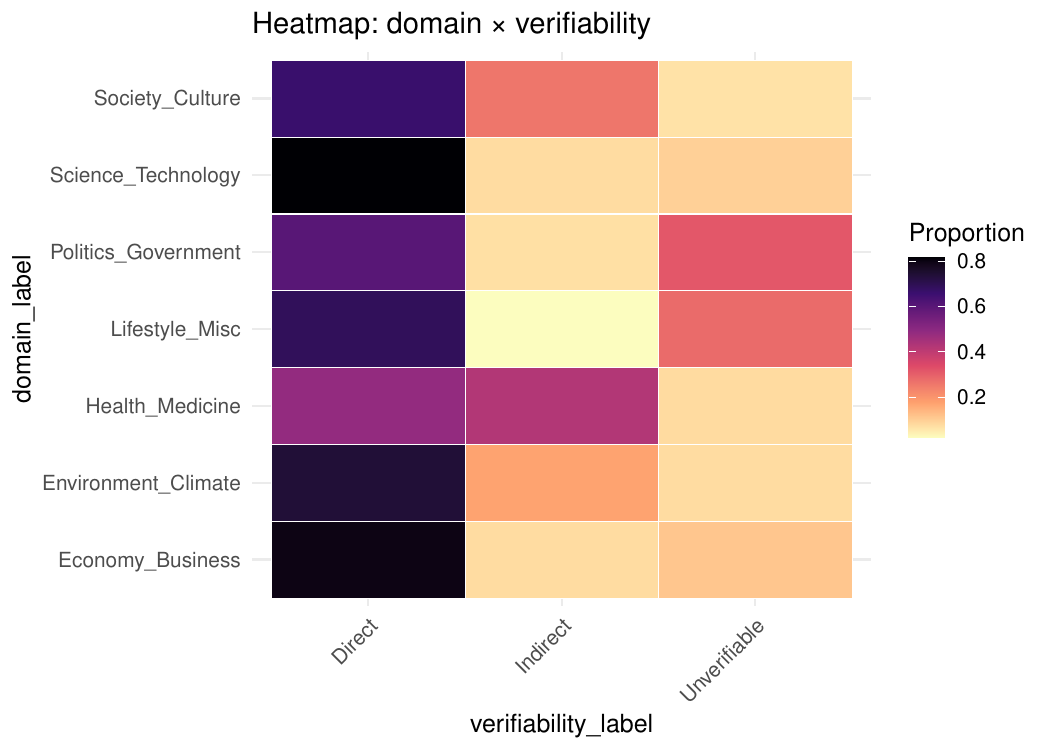}
  \includegraphics[width=\linewidth]{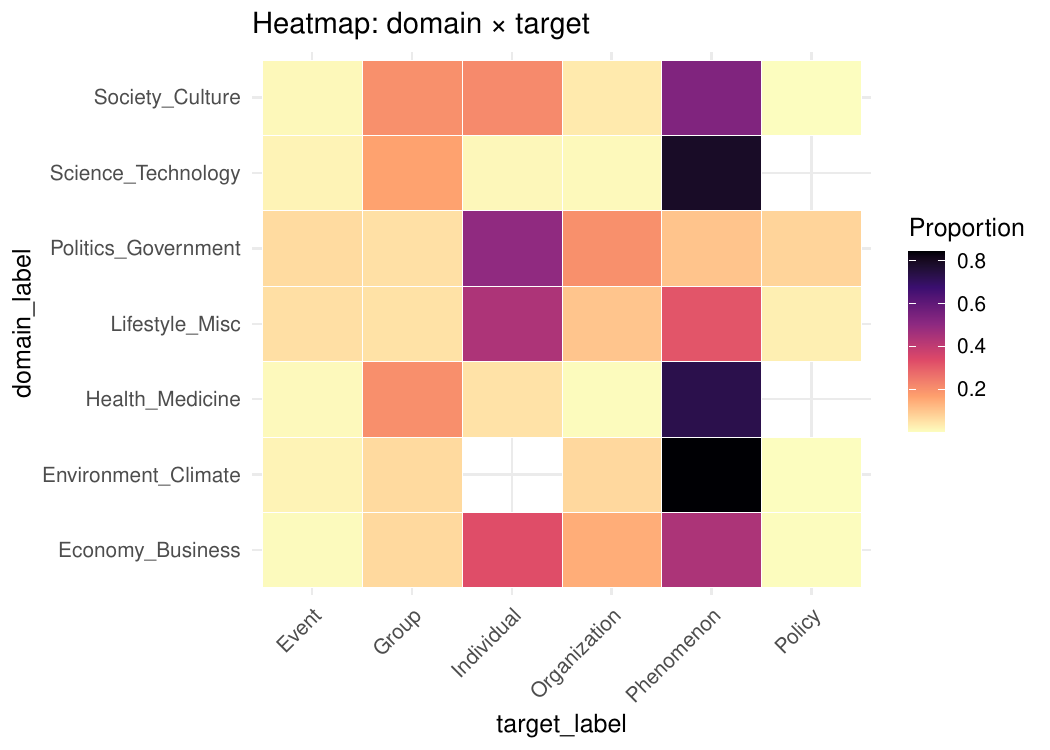}
\end{minipage}
\hfill
\begin{minipage}{0.45\linewidth}
  \includegraphics[width=\linewidth]{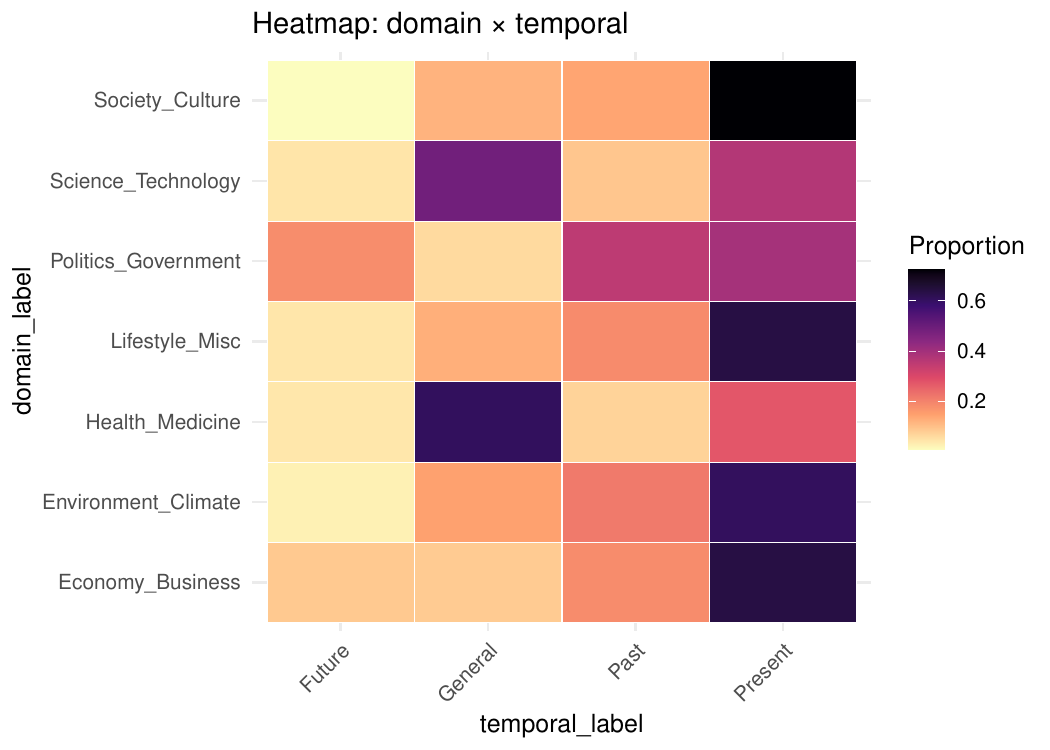}
  \includegraphics[width=\linewidth]{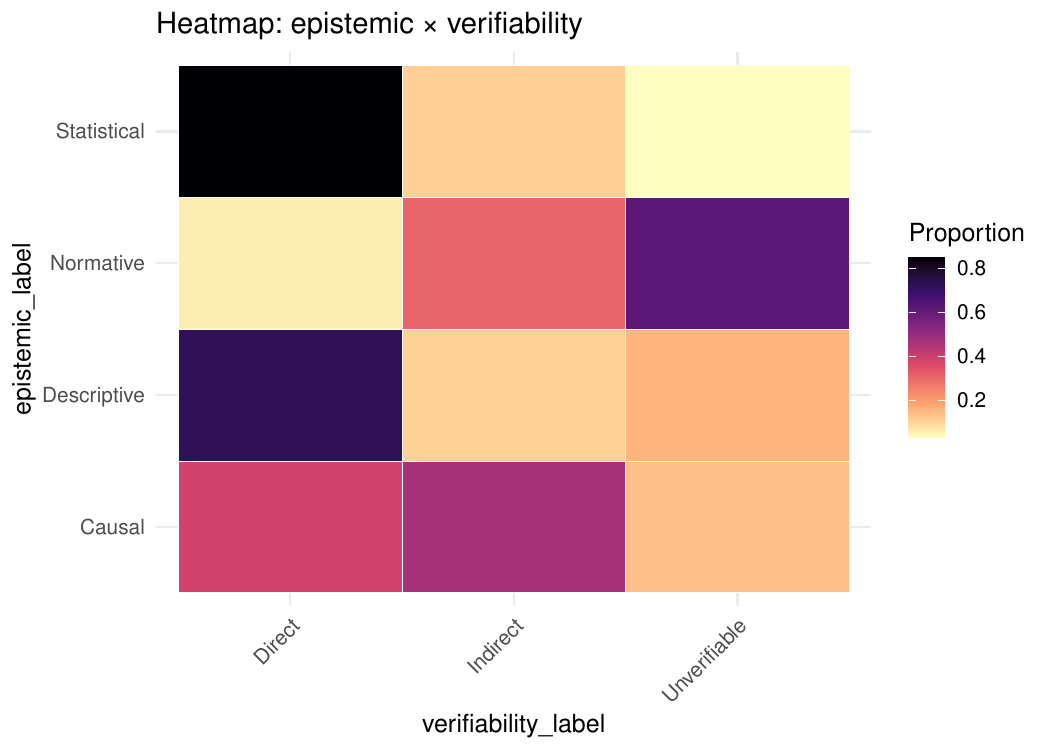}
\end{minipage}
\caption{Cross-tab associations between \textit{domain} $x$ \textit{epistemic type}, \textit{domain} $x$ \textit{verifiability}, \textit{domain} $x$ \textit{target}, and \textit{epistemic type} $x$ \textit{target}.}
\end{figure}

\begin{figure}[H]
\centering
\begin{minipage}{0.45\linewidth}
  \includegraphics[width=\linewidth]{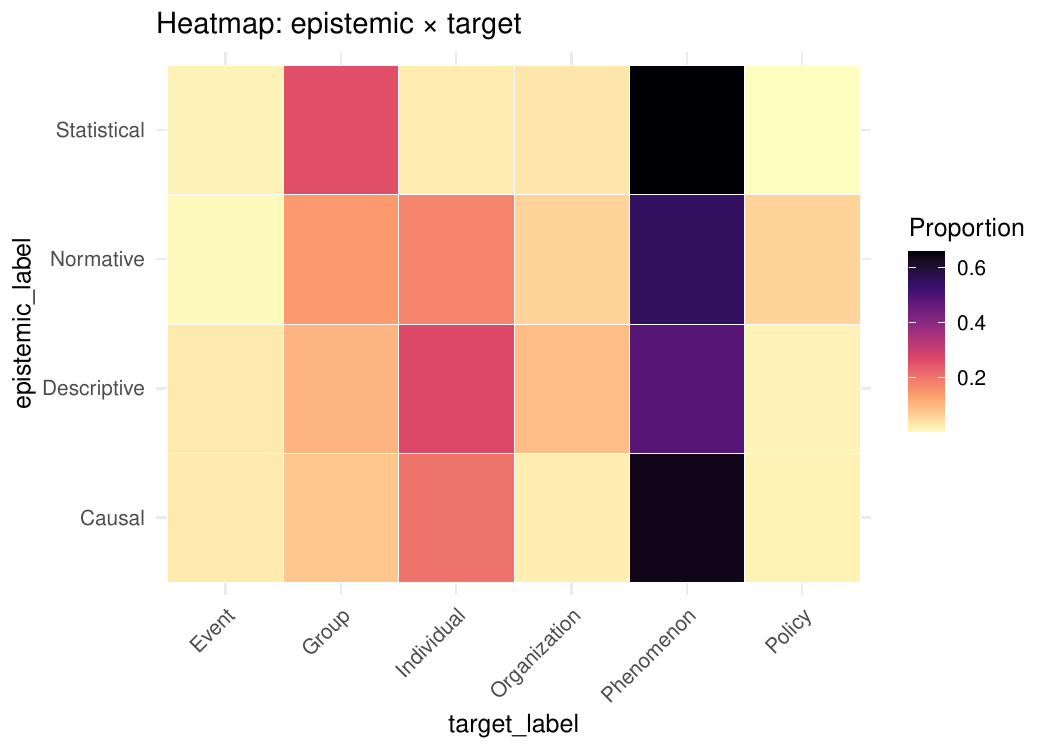}
  \includegraphics[width=\linewidth]{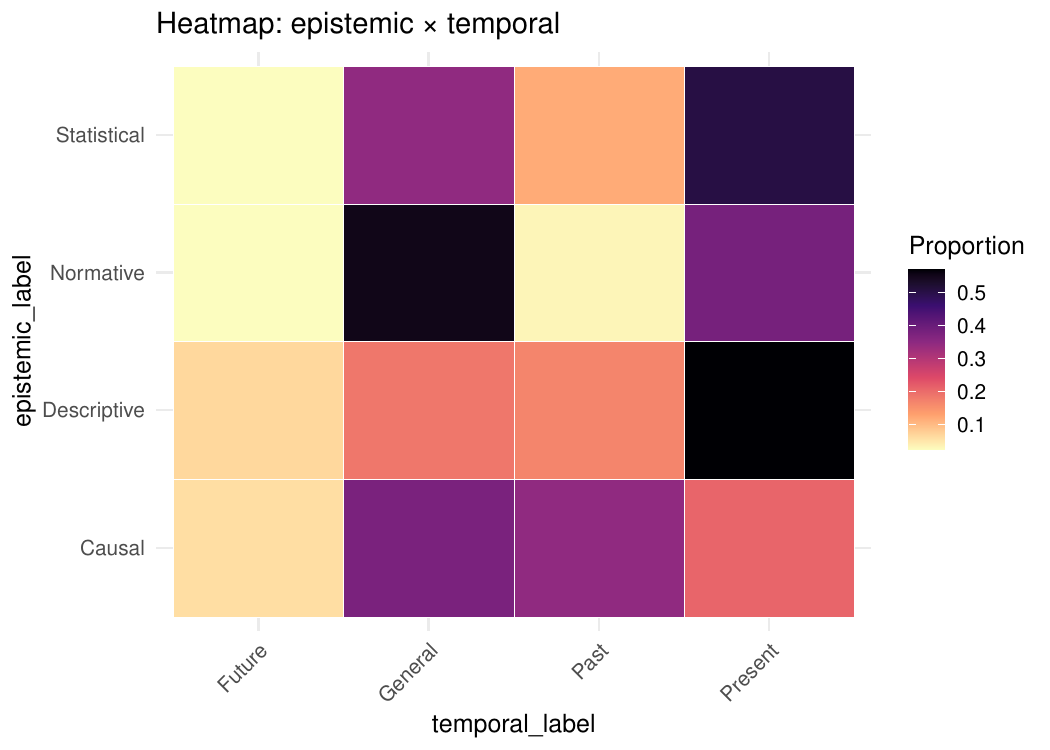}
  \includegraphics[width=\linewidth]{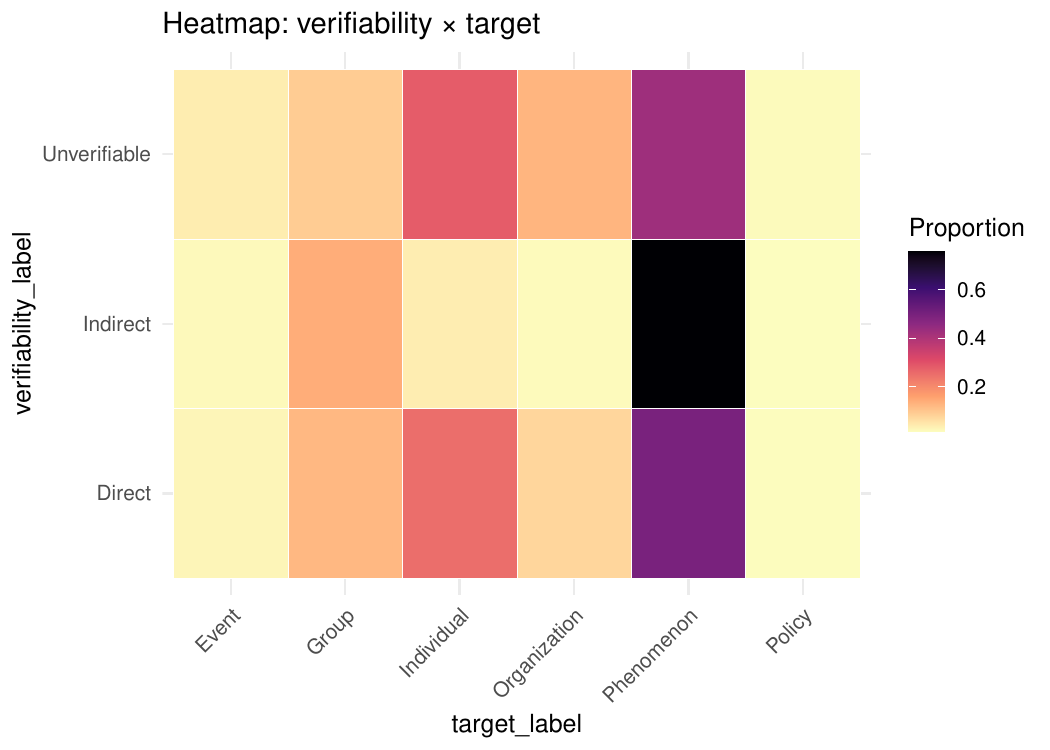}
\end{minipage}
\hfill
\begin{minipage}{0.45\linewidth}
  \includegraphics[width=\linewidth]{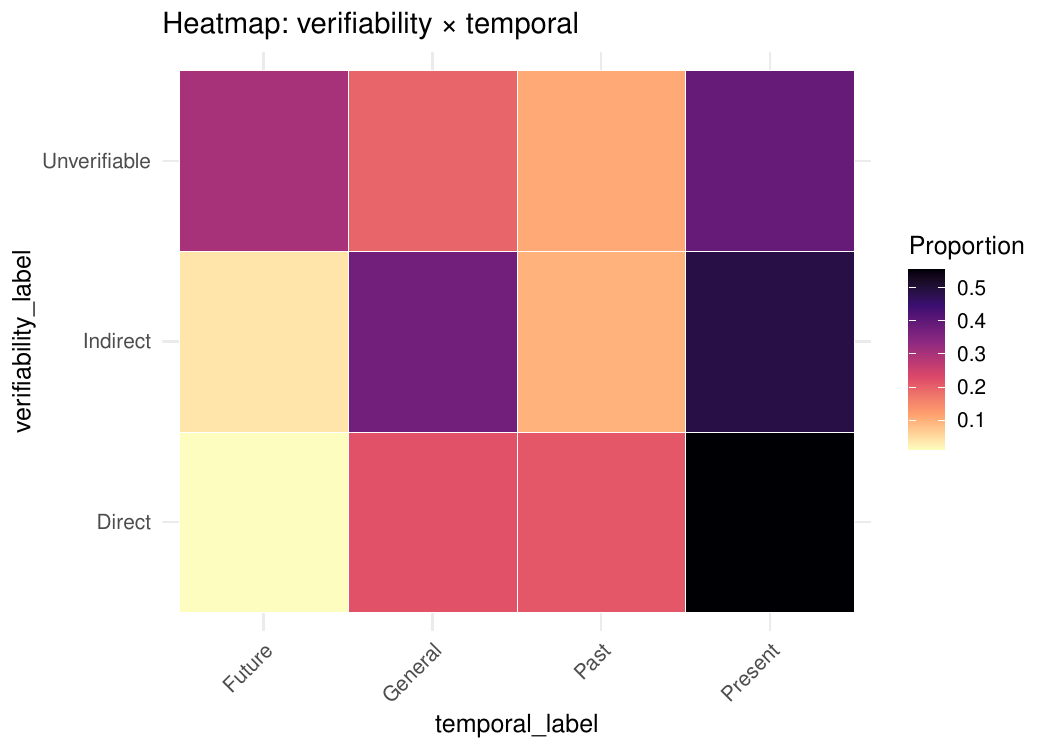}
  \includegraphics[width=\linewidth]{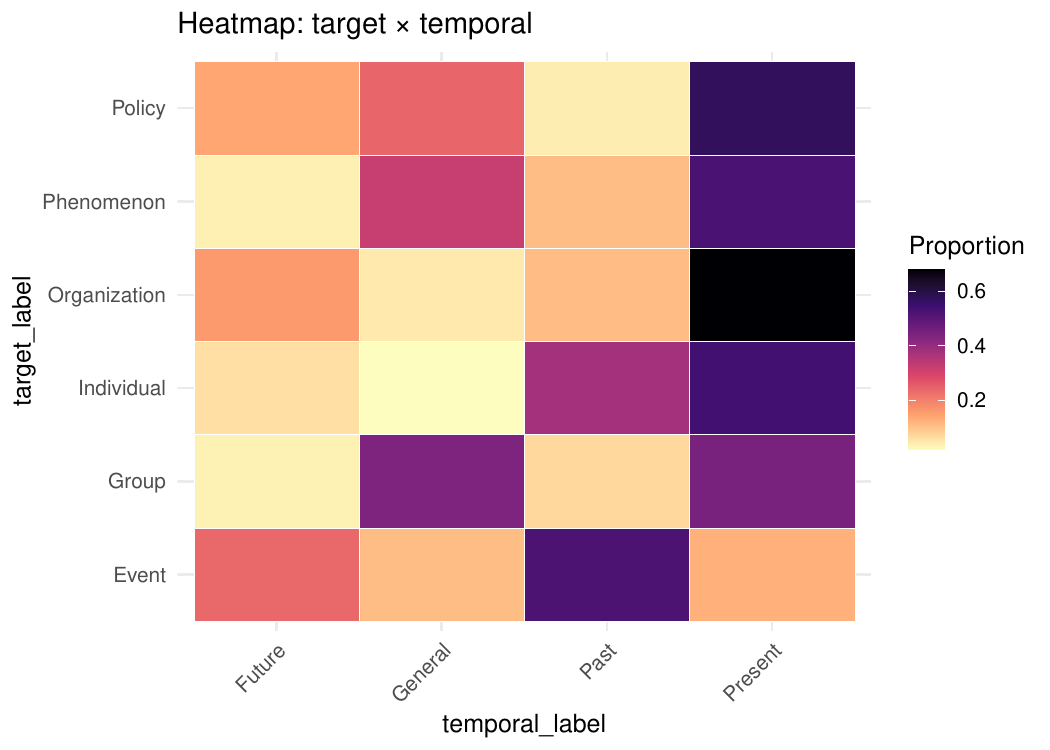}
\end{minipage}
\caption{Cross-tab associations between \textit{epistemic type} $x$ \textit{target}, \textit{epistemic type} $x$ \textit{temporal}, \textit{verifiability} $x$ \textit{target}, \textit{verifiability} $x$ \textit{temporal}, and \textit{target} $x$ \textit{temporal}.}
\end{figure}

\section{Comparison with FEVER}
\label{app:fever}

\begin{table}[htbp]
\caption{Comparison of domain classification proportions}
\centering
\begin{tabular}[t]{llll}
\toprule
Label & User dataset (\%) & FEVER (\%) & $\Delta$ (pp)\\
\midrule
Lifestyle\_Misc & 16.1 & 63.2 & -47.0\\
Science\_Technology & 22.3 & 3.1 & +19.2\\
Health\_Medicine & 13.9 & 1.7 & +12.2\\
Politics\_Government & 18.6 & 7.2 & +11.5\\
Environment\_Climate & 8.5 & 1.3 & +7.2\\
Society\_Culture & 12.8 & 19.8 & -7.0\\
Economy\_Business & 7.7 & 3.7 & +4.0\\
\bottomrule
\end{tabular}
\end{table}

\begin{table}[htbp]
\caption{Comparison of epistemic classification proportions between datasets.}
\centering
\begin{tabular}[t]{llll}
\toprule
Label & User dataset (\%) & FEVER (\%) & $\Delta$ (pp)\\
\midrule
Descriptive & 72.1 & 95.9 & -23.8\\
Causal & 11.8 & 1.4 & +10.5\\
Statistical & 11.5 & 2.8 & +8.8\\
Normative & 4.5 & 0.0 & +4.5\\
\bottomrule
\end{tabular}
\end{table}

\begin{table}[htbp]
\caption{Comparison of verifiability classification proportions between datasets.}
\centering
\begin{tabular}[t]{llll}
\toprule
Label & User dataset (\%) & FEVER (\%) & $\Delta$ (pp)\\
\midrule
Direct & 64.3 & 95.5 & -31.2\\
Unverifiable & 21.4 & 3.1 & +18.3\\
Indirect & 14.3 & 1.4 & +12.9\\
\bottomrule
\end{tabular}
\end{table}

\begin{table}[htbp]
\caption{Comparison of target classification proportions between datasets.}
\centering
\begin{tabular}[t]{llll}
\toprule
Label & User dataset (\%) & FEVER (\%) & $\Delta$ (pp)\\
\midrule
Individual & 19.2 & 53.5 & -34.3\\
Phenomenon & 50.5 & 24.2 & +26.3\\
Group & 15.8 & 4.8 & +11.0\\
Event & 4.7 & 7.8 & -3.0\\
Policy & 2.3 & 0.2 & +2.1\\
Organization & 7.5 & 9.5 & -2.0\\
\bottomrule
\end{tabular}
\end{table}

\begin{table}[htbp]
\caption{Comparison of temporal classification proportions between datasets.}
\centering
\begin{tabular}[t]{llll}
\toprule
Label & User dataset (\%) & FEVER (\%) & $\Delta$ (pp)\\
\midrule
Past & 18.6 & 55.6 & -37.0\\
General & 29.9 & 7.6 & +22.2\\
Future & 8.7 & 0.5 & +8.2\\
Present & 42.8 & 36.3 & +6.5\\
\bottomrule
\end{tabular}
\end{table}

\end{document}